\documentclass[journal]{IEEEtran}
\ifCLASSINFOpdf
\else
\fi
\usepackage{textcomp,marvosym}
\usepackage{lineno,hyperref}

\usepackage{cite}

\usepackage[utf8]{inputenc}
\usepackage[T1]{fontenc}
\usepackage{lineno,hyperref}
\usepackage{bm}
\usepackage{amssymb}
\usepackage{pdflscape}
\usepackage{algorithm}
\usepackage{algpseudocode}
\usepackage{lipsum}
\usepackage{algpseudocode}
\usepackage{float}
\floatstyle{boxed} 
\usepackage{latexsym} 
\usepackage{graphics} 
\usepackage{epsfig} 
\usepackage{times} 
\usepackage{amsmath} 
\usepackage{amssymb}  
\usepackage{graphicx}
\usepackage{amsmath,amssymb,amsfonts}
\usepackage{caption}
\usepackage{booktabs}
\usepackage{epstopdf}
\usepackage{color}
\usepackage{subfigure}
\usepackage{hyperref}
\usepackage{multirow,tabularx}
\usepackage{picins}
\usepackage{changepage}
\usepackage{textcomp,marvosym}
\usepackage{nameref,hyperref}
\usepackage{microtype}
\usepackage{array}

\begin{document}
%
\title{\emph{4S-DT}: Self Supervised Super Sample Decomposition for Transfer learning with application to COVID-19 detection}
%
%
%

\author{Asmaa Abbas, Mohammed M. Abdelsamea, and Mohamed Medhat Gaber
\thanks{A. Abbas is with Mathematics Department, University of Assiut, Assiut, Egypt}
\thanks{M. Abdelsamea is with School of Computing and Digital Technology, Birmingham City University, Birmingham, UK and Mathematics Department, University of Assiut, Assiut, Egypt}
\thanks{M. Gaber is with School of Computing and Digital Technology, Birmingham City University, Birmingham, UK}
}

%
%

\markboth{Journal of \LaTeX\ Class Files,~Vol.~14, No.~8, August~2015}%
{Shell \MakeLowercase{\textit{et al.}}: Bare Demo of IEEEtran.cls for IEEE Journals}
%



\maketitle

\begin{abstract}
Due to the high availability of large-scale annotated image datasets, knowledge transfer from pre-trained models showed outstanding performance in medical image classification. However, building a robust image classification model for datasets with data irregularity or imbalanced classes can be a very challenging task, especially in the medical imaging domain. In this paper, we propose a novel deep convolutional neural network, we called Self Supervised Super Sample Decomposition for Transfer learning (\emph{4S-DT}) model. \emph{4S-DT} encourages a coarse-to-fine transfer learning from large-scale image recognition tasks to a specific chest X-ray image classification task using a generic self-supervised sample decomposition approach. Our main contribution is a novel self-supervised learning mechanism guided by a super sample decomposition of unlabelled chest X-ray images. \emph{4S-DT} helps in improving the robustness of knowledge transformation via a downstream learning strategy with a class-decomposition layer to simplify the local structure of the data. \emph{4S-DT} can deal with any irregularities in the image dataset by investigating its class boundaries using a downstream class-decomposition mechanism. We used 50,000 unlabelled chest X-ray images to achieve our coarse-to-fine transfer learning with an application to COVID-19 detection, as an exemplar. \emph{4S-DT} has achieved a high accuracy of $99.8\%$ (95$\%$ CI: 99.44 \%, 99.98\%) in the detection of COVID-19 cases on a large dataset and an accuracy of $97.54\%$ (95$\%$ CI: 96.22\%, 98.91\%) on an extended test set enriched by augmented images of a small dataset, out of which all real COVID-19 cases were detected, which was the highest accuracy obtained when compared to other methods.

\end{abstract}

\begin{IEEEkeywords}
self-supervision; chest X-ray image classification; transfer learning; data irregularities; convolutional neural network.
\end{IEEEkeywords}

%
\IEEEpeerreviewmaketitle

\section{Introduction}

\begin{figure}[hbt!]
    \centering
        \includegraphics[scale=0.4]{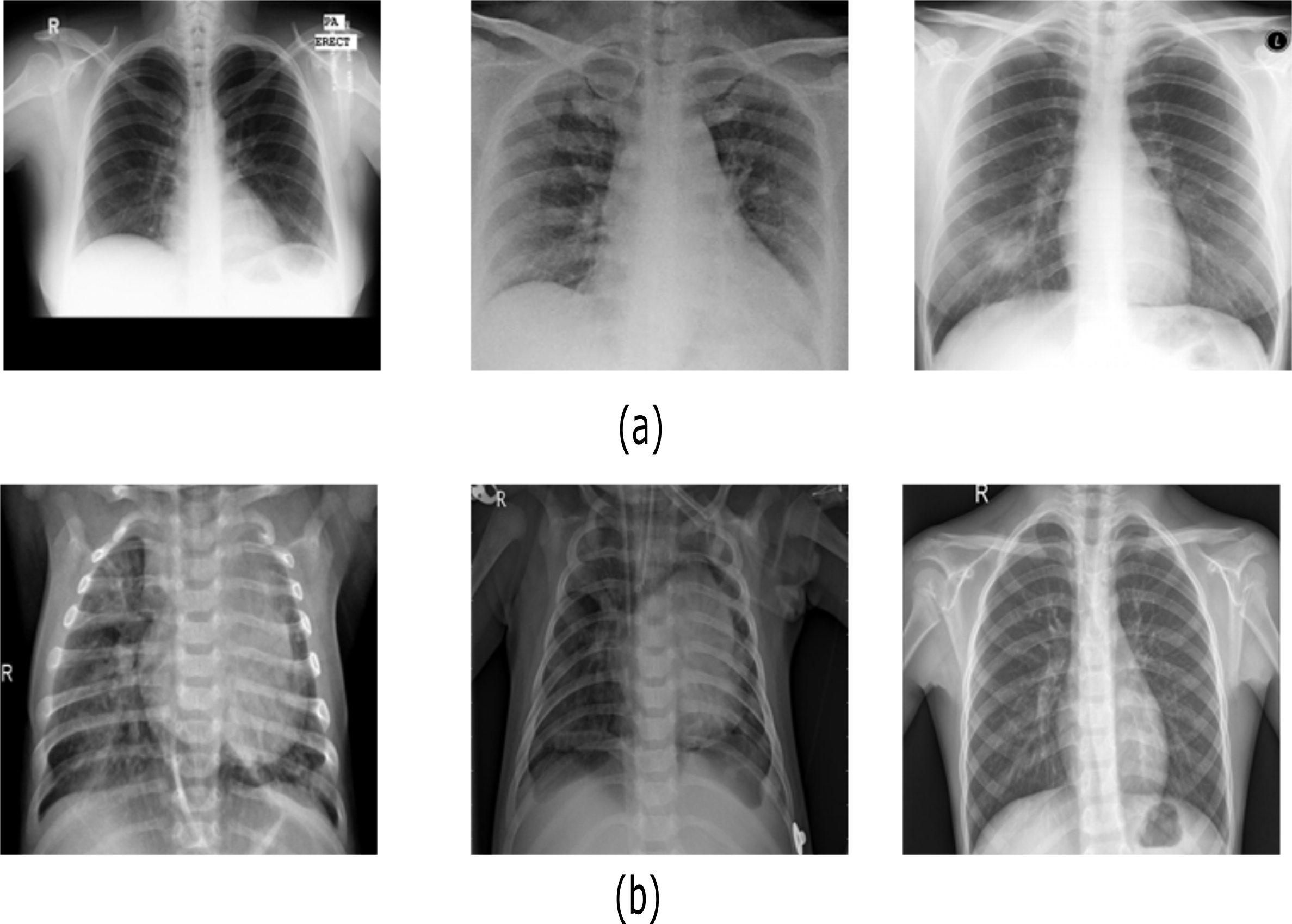}
        \caption{Examples of a) labelled chest X-ray images (from left to right: normal, COVID-19, and SARS images), and b) unlabelled chest X-ray images used in this work for self-supervision learning.}
        \label{Figimage}
    \end{figure}

Diagnosis of COVID-19 is associated with the symptoms of pneumonia and chest X-ray tests \cite{shi2020radiological}. Chest X-ray is the essential imaging technique that plays an important role in the diagnosis of COVID-19 disease. Fig. \ref{Figimage} shows examples of a) a normal chest X-ray, a positive one with COVID-19, a positive image with the severe acute respiratory syndrome (SARS), and b) some examples of other unlabelled chest X-ray images used in this work. 

Several statistical machine learning methods have been previously used for automatic classification of digitised lung images \cite{dandil2014artificial, kuruvilla2014lung}. For instance, in \cite{manikandan2016lung}, a small set of three statistical features were calculated from lung texture to distinguish between malignant and benign lung nodules using a Support Vector Machine \emph{SVM} classifier. A statistical co-occurrence matrix method was used with Backpropagation Network \cite{sangamithraa2016lung} to classify samples from being normal or cancerous. With the high availability of enough annotated image data, deep learning approaches \cite{PESCE201926, XIE2019237, 8639200} usually provide a superiority performance over the statistical machine learning approaches. Convolutional Neural Networks (\emph{CNN}) is one of the most commonly used deep learning approaches with superior achievements in the medical imaging domain \cite{lecun2015deep}. The primary success of \emph{CNN} is due to its capability to learn local features automatically from domain-specific images, unlike the statistical machine learning methods. One of the popular strategies for training a \emph{CNN} model is to transfer learned knowledge from a pre-trained network that fulfilled one generic task into a new specific task \cite{pan2009survey}. Transfer learning is faster and easy to apply without the need for a huge annotated dataset for training; therefore many scientists tend to adopt this strategy especially with medical imaging. Transfer learning can be accomplished with three main scenarios \cite{li2014medical}: a) ``shallow tuning'', which adapts only the classification layer in a way to cope with the new task, and freezes the weights of the remaining layers without updating; b) ``deep tuning'' which aims to retrain all the weights of the adopted pre-trained network from end-to-end; and (c) ``fine-tuning'' that aims to gradually train layers by tuning the learning parameters until a significant performance boost is achieved. Transfer knowledge via fine-tuning scenario demonstrated outstanding performance in chest X-ray and computed tomography image classification \cite{JOYSEEREE2019172, gao2018holistic}.


The emergence of COVID-19 as a pandemic disease dictated the need for faster detection methods to contain the spread of the virus. As aforementioned, chest X-ray imaging comes in as a promising solution, particularly when combined with an effective machine learning model. In addition to data irregularities that can be dealt with through class decomposition, scarcity of data, especially in the early months of the pandemic, made it hard to realise the adoption of chest X-ray images as a means for detection. On the other hand, self-supervised learning is being popularised recently to address the expensive labelling of data acquired at an unprecedented rate. In self-supervised learning, unlabelled data is used for feature learning by assigning each example a pseudo label. In the case of convolutional neural networks (CNN) applied on image data, each image is assigned a pseudo label, and CNN is trained to extract visual features of the data. The training of a CNN by pseudo labelled images as input is called pretext task learning. While the training of the produced CNN from the pretext training using labelled data is called downstream task training. Inherently such a pipeline allows for effective utilisation of large unlabelled data sets. The success of the pretext task learning relies on pseudo labelling methods. In \cite{jing2020self}, four categories of methods were identified. Context-based image feature learning by means of context similarity has demonstrated a particularly effective pseudo labelling mechanism. DeepCluster \cite{caron2018deep} is the state-of-the-art method under this category. DeepCluster is a super sample decomposition method that generates pseudo labels through the clustering of CNN features. Sample decomposition is the process of applying clustering on the whole training set as a step for improving supervised learning performance \cite{rokach2005improving}. When the clustering is performed on a larger data sample, we refer to this process as a super sample decomposition. However, we argue that the coupling of the pretext task and the pseudo labelling can limit the effectiveness of the pretext task in the self-supervised learning process. In our proposed super sample decomposition, the pretext task training uses cluster assignments as pseudo labels, where the clustering process is decoupled from the pretext training. We propose the clustering of encoded images through an auto-encoder neural network, allowing flexibility of utilising different features and clustering methods, as appropriate. We argue that this can be most effective in medical image classification, evident by the experimentally validated use of class decomposition for transfer learning in a method coined as DeTraC \cite{abbas2020detrac}.            

In this paper, we propose a novel deep convolutional neural network, we term Self Supervised Super Sample Decomposition for Transfer learning (\emph{4S-DT}) model for the detection of COVID-19 cases \footnote{The developed code is available at https://github.com/asmaa4may/4S-DT.}. \emph{4S-DT} has been designed in a way to encourage a coarse-to-fine transfer learning based on a self-supervised sample decomposition approach. \emph{4S-DT} can deal with any irregularities in the data distribution and the limited availability of training samples in some classes. The contributions of this paper can be summarised as follows. We provide

\begin{itemize}
\item a novel mechanism for self-supervised sample decomposition using a large set of unlabelled chest X-ray images for a pretext training task;
\item a generic coarse-to-fine transfer learning strategy to gradually improve the robustness of knowledge transformation from large-scale image recognition tasks to a specific chest X-ray image classification task;
\item a downstream class-decomposition layer in the downstream training phase to cope with any irregularities in the data distribution and simplify its local structure; and
\item a thorough experimental study on COVID-19 detection, pushing the boundaries of state-of-the-art techniques in terms of accuracy, and robustness of the proposed model.  
\end{itemize}

The paper is organised as follow. In Section \ref{relatedwork}, we review the state-of-the-art methods for COVID-19 detection. Section \ref{methods} discusses the main components of our proposed \emph{4S-DT} model. Section \ref{results} describes our experiments on several chest X-ray images collected from different hospitals. In Section \ref{discussion}, we discuss our findings and conclude the work.  

\section{Previous work on COVID-19 detection from chest X-ray}
\label{relatedwork}

In February 2020, the World Health Organisation (WHO) has declared that a new virus called COVID-19 has started to spread aggressively in several countries \cite{world2020coronavirus}. Diagnosis of COVID-19 is typically associated with pneumonia-like symptoms, which can be revealed by both genetic and imaging tests. Fast detection of the virus will directly contribute to managing and controlling its spread. 
Imaging tests, especially chest X-ray, can provide fast detection of COVID-19 cases. The historical conception of medical image diagnostic systems has been comprehensively explored through an enormous number of approaches ranging from statistical machine learning to deep learning. 
A convolutional neural network is one of the most effective approaches in the diagnosis of lung diseases including COVID-19 directly from chest X-ray images. Several recent reviews have been carried out to highlight significant contributions to the detection of COVID-19 \cite{shi2020review, 9079648, li2020artificial}. For instance, in \cite{song2020deep}, a modified version of ResNet-50 pre-trained \emph{CNN} model has been used to classify \emph{CT} images into three classes: healthy, COVID-19 and bacterial pneumonia. In \cite{wang2020covidnet}, a \emph{CNN} model, called COVID-Net, based on transfer learning was used to classify chest X-ray images into four classes: normal, bacterial infection, non-COVID, and COVID-19 viral infection. In \cite{9097297}, a weakly-supervised approach has been proposed using 3D chest CT volumes for COVID-19 detection and lesion localisation relying on ground truth masks obtained by an unsupervised lung segmentation method and a 3D ResNet pre-trained model. In \cite{apostolopoulos2020covid}, a dataset of chest X-ray images from patients with pneumonia, confirmed COVID-19 disease, and normal incidents, was used to evaluate the performance of the state-of-the-art \emph{CNN} models based on transfer learning. The study suggested that transfer learning can provide important biomarkers for the detection of  COVID-19 cases. It has been experimentally demonstrated that transfer learning can provide a robust solution to cope with the limited availability of training samples from confirmed COVID-19 cases \cite{9090149}.

In \cite{chen2019self}, self-supervised learning using context distortion is applied for classification, segmentation, and localisation in different medical imaging problems. When used in classification, the method was applied for scan plane detection in fetal 2D ultrasound images, showing classification improvement in some settings. However, we argue that our proposed method is more effective in image segmentation and localisation, because context distortion is able to generate localised features, instead of global image features that can be more effective for classification tasks. 

Having reviewed the related work, it is evident that despite the great success of deep learning in the detection of COVID-19 cases from chest X-ray images, data scarcity and irregularities have not been explored. It is common in medical imaging in particular that datasets exhibit different types of irregularities (e.g. overlapping classes with imbalance problems) that affect the resulting accuracy of deep learning models. With the unfolding of COVID-19, chest X-ray images are rather scarce. Thus, this work focuses on coping with data irregularities through class decomposition, and data scarcity through super sample decomposition, as detailed in the following section.

\section{\emph{4S-DT} model}
\label{methods}

\begin{figure*}[]
    \centering
        \includegraphics[scale=0.18]{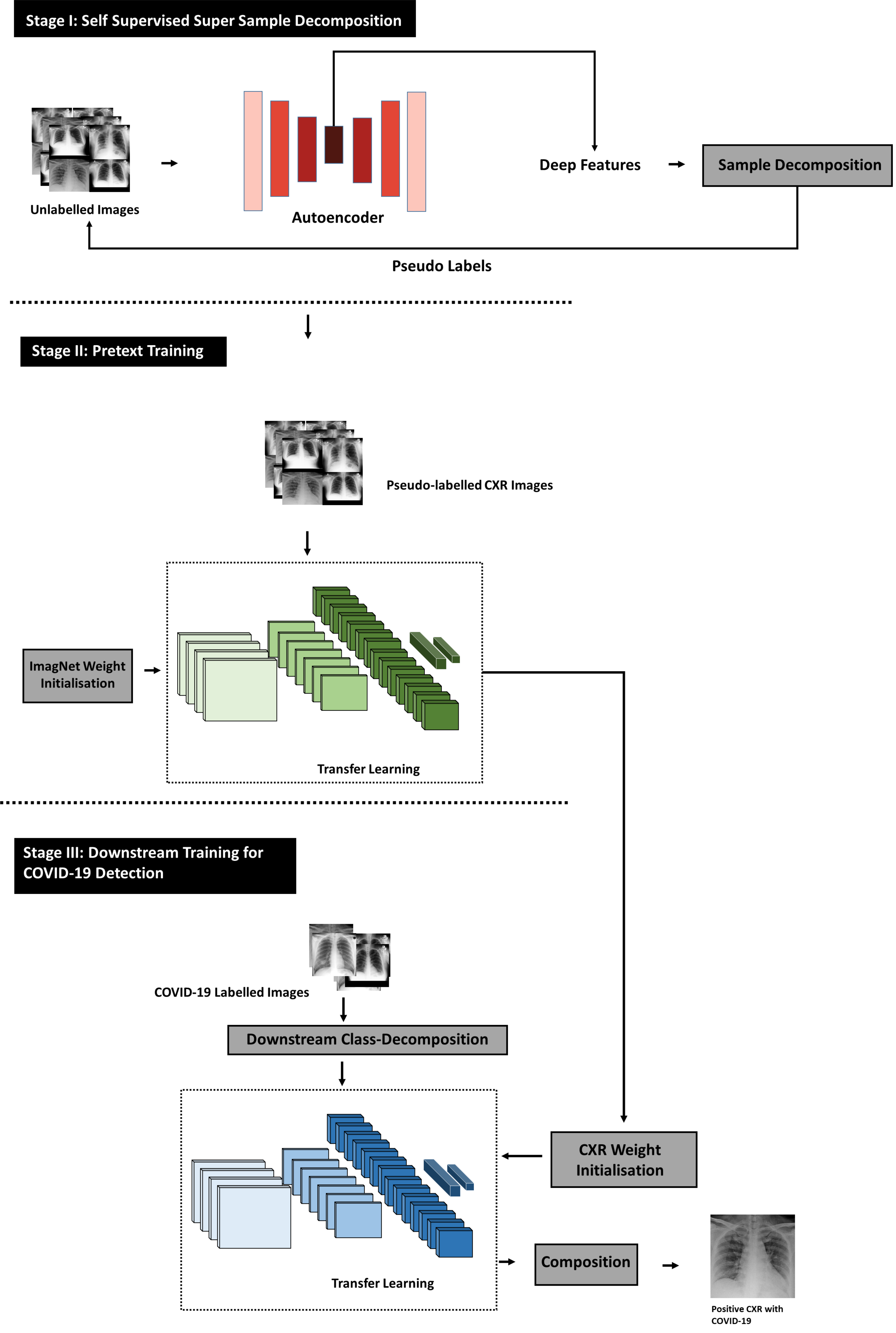}
        \caption{Graphical representation of \emph{4S-DT} model.}
        \label{Figproposed}
    \end{figure*}

This section describes, in sufficient details, our proposed deep convolutional neural network, Self Supervised Super Sample Decomposition for Transfer learning (\emph{4S-DT}) model for detecting COVID-19 cases from chest X-ray images. Starting with an overview of the architecture through to the different components of the model, the section discusses the workflow and formalises the method.
\emph{4S-DT} model consists of three training phases, see Fig. \ref{Figproposed}. In the first phase, we train an autoencoder model to extract deep local features from each sample in a super large set of unlabelled generic chest X-ray images. Then we adapted a sample decomposition mechanism to create pseudo labels for the generic chest X-ray images. In the second phase, we use the pseudo labels to achieve a coarse transfer learning using an ImageNet pre-trained \emph{CNN} model for the classification of pseudo-labelled chest X-ray images (as a pretext training task), resulting in a chest X-ray-related convolutional features.  
In the last phase, we use trained convolutional features to achieve downstream training. The downstream training task is more task-specific by adapting a fine transfer learning from chest X-ray recognition to COVID-19 detection. In this stage, we also adapt a class-decomposition layer to simplify the local structure of the image data distribution, where a sophisticated gradient descent optimisation method is used. Finally, we apply a class-composition to refine the final classification of the images.

\subsection{Super sample decomposition}

Given a set of unlabelled images $X=\{x^{1},x^{2},...,x^{n'}\}$, our super sample decomposition component aims to find and use pseudo labels during the pretext training task of \emph{4S-DT}. To this end, an autoencoder (\emph{AE}) is first used to extract deep features associated to each image. For each input image $x$, the representation vector $ h^d$ and the reconstructed image $\hat{x}$ can be defined as 
\begin{align} h^{d} = f(W^{(1)}{x}+b^{(1)}) \\[3pt] \hat {x} = f( {W}^{(2)}{h}^{d}+{b}^{(2)}) {}\end{align}

where $W^{(1)}$ and $W^{(2)}$ are the weight matrices, $b^{(1)}$ and $b^{(2)}$ are the bias vectors, and $f$ is the active function. The reconstruction error $L(x,\hat{x})$ between $\hat{x}$ and $x$ is defined as

\begin{align} L(x,\hat {x})=&\frac {1}{2}\left \|x-\hat {x}\right \|^{2} {}\end{align}

The overall cost function of the $n'$ unlabelled images,$E_{AE}({ {W}},{ {b}})$, can be defined as 

\begin{align} E_{AE}({ {W}},{ {b}})=&\left [{ {\frac {1}{n'}\sum \limits _{i=1}^{n'} {L({ {x}}^{i},{\hat { {x}}}^{i})} } }\right ]+\frac {\lambda }{2}\sum \limits _{l=1}^{n_{l} -1} {\sum \limits _{i=1}^{s_{l} } {\sum \limits _{j=1}^{s_{l+1} } {({ {W}}_{ji}^{(l)} )^{2}} } }\notag \\ {}\end{align}

where the first term denotes the reconstruction error of the whole datasets, and the second term is the regularisation weight penalty term, which aims to prevent over-fitting by restraining the magnitude of the weights. $\lambda$ is the weight decay parameter, $n_l$ is the layer number of the network, $s_l$ denotes the neuron number in layer $l$ , and $W^{(l)}_{ji}$ is the connecting weight between neuron $i$ in layer $l+1$ and neuron $j$ in layer $l$.




Once the training of the \emph{AE} has been accomplished, Density-Based Spatial Clustering of Applications with Noise (\emph{DBSCAN}) is used to cluster the image data distribution $X$ into a number of classes $c$ based on the extracted features $h^{d}$. \emph{DBSCAN} is an unsupervised clustering algorithm, which is a considerably representative density-based clustering algorithm that defines clusters as the largest set of points connected by density.

Let the image dataset $X$ be mapped into a low-dimensional feature space denoted by $H \in R^{n'\times d}$, where $H=(h_1,h_2,...,h_{n'})$. An image $x^i$ (represented by $h_i$) is density-connected to image $x^j$ (represented by $h_j$) with respect to $Eps$ (i.e. neighbourhood radius) and $MinPts$ (i.e. the minimum number of objects within the neighbourhood radius of core object) if there exists a core object $x^k$ such that both $x^i$ and $x^j$ are directly density-reachable from $x^k$ with respect to $Eps$ and $MinPts$. An image $x^i$ is directly density-reachable from an image $x^j$ if $x^i$ is within the $Eps$-neighbourhood of $N_{Eps}(x^j)$, and $x^j$ is a core object, where $Eps$-neighbourhood can be defined as

\begin{eqnarray}
N_{Eps}(x_j)=\{x_i \in X | dis(x_i,x_j) \leq Eps\}.
\end{eqnarray}

\emph{DBSCAN} results in $C$ clusters, where each cluster is constructed by maximising the density reachability relationship among images of the same cluster. The $C$ cluster labels will be assigned to the $n'$ unlabelled images and will be presented as pseudo labels for the pretext training task and hence the downstream training task. The pseudo-labelled image dataset can then be defined as $X'= \{ (x^i,y^c) | c \in C \}$.


\subsection{Pretext training}
With the high availability of large-scale annotated image datasets, the chance for the different classes to be well-represented is high. Therefore, the learned in-between class-boundaries are most likely to be generic enough to new samples. On the other hand, with the limited availability of annotated medical image data, especially when some classes are suffering more compared to others in terms of the size and representation, the generalisation error might increase. This is because there might be a miscalibration between the minority and majority classes. Large-scale annotated image datasets (such as ImageNet) provide effective solutions to such a challenge via transfer learning where tens of millions of parameters (of \emph{CNN} architectures) are required to be trained.

A shallow-tuning mode was used during the adaptation and training of an ImageNet pre-trained \emph{CNN} model using the collected chest X-ray image dataset. We used the off-the-shelf \emph{CNN} features of pre-trained models on ImageNet (where the training is accomplished only on the final classification layer) to construct the image feature space. 

Mini-batch of stochastic gradient descent (\emph{mSGD}) was used to minimise the categorical cross entropy loss function, $E_{coarse}(\cdot)$ 
\begin{eqnarray}
E_{coarse}\left ( y^{c},z'(x^{j}, W') \right ) & = & -\sum_{c=1}^{C} y^{c} \ln {z'\left(x^{j}, W'\right )},
\end{eqnarray}

where \( x^{j} \) is the set of self-labelled images in the training, \( y^{c} \) is their associated self labels while $z'\left(x^{j}, W'\right )$ is the predicted output from a softmax function, where $W'$ is the converged weight matrix associated to the ImageNet pre-trained model (i.e. we used $W'$ of ImageNet pre-trained CNN model for weight initialisation to achieve a coarse transfer learning).

\subsection{Downstream training}

A fine-tuning mode was used during the adaptation of \emph{ResNet} model using feature maps from the coarse transfer learning stage. However, due to the high dimensionality associated with the images, we applied \emph{PCA} to project the high-dimension feature space into a lower-dimension, where highly correlated features were ignored. This step is important for the downstream class-decomposition process in the downstream training phase to produce more homogeneous classes, reduce the memory requirements, and improve the efficiency of the framework.

        Now assume that our feature space (\emph{PCA}'s output) is represented by a 2-D matrix (denoted as dataset $A$), and \( \mathbf{L} \)  is a class category. \(  A \) and \(  \mathbf{L} \) can be rewritten as
    \begin{equation}
      A= \left[ \begin{matrix}
        a_{11}  &  a_{11}  &   \ldots  ~~~~~ a_{1m}\\
        a_{21}  &  a_{22}  &   \ldots ~~~~~~ a_{2m}\\
         \vdots   &   \vdots   &   \vdots ~~~~~~~~~~~  \vdots \\
        a_{n1}  &  a_{n2}  &   \ldots ~~~ a_{nm}\\
        \end{matrix}
         \right]  ,   \mathbf{L}= \left\{ l_{1}, l_{2}, \ldots ,l_{c'} \right\},
    \end{equation}

where $n$ is the number of images,  \(  m \) is the number of features, and $c'$  is the number of classes. For downstream class-decomposition, we used $k$-means clustering \cite{wu2008top} to further divide each class into homogeneous sub-classes (or clusters), where each pattern in the original class \(  \mathbf{L}  \)  is assigned to a class label associated with the nearest centroid based on the squared euclidean distance (\emph{SED}):
    \begin{equation}
     SED= \sum _{j=1}^{k} \sum _{i=1}^{n}\parallel a_{i}^{ \left( j \right) }-c_{j}\parallel, 
    \end{equation}

where centroids are denoted as  \( c_{j} \).
Once the clustering is accomplished, each class in  \( \mathbf{L} \) will further be divided into $k$ subclasses, resulting in a new dataset (denoted as dataset $B$). 
Accordingly, the relationship between dataset $A$ and $B$ can be mathematically described as:
\begin{equation}
 A = ( A | \mathbf{L} ) ~ \mapsto~ B= (B | \mathbf{C'} )
\end{equation}

where the number of instances in  $A$  is equal to $B$ while $\mathbf{C'}$ encodes the new labels of the subclasses (e.g. $\mathbf{C'} =\{l_{11}, l_{12}, \dots, l_{1k}, l_{21}, l_{22}, \dots, l_{2k}, \dots l_{c'k} \}$). 

For transfer learning, we used ResNet \cite{he2016deep} model, which showed excellent performance with only 18 layers. Here we consider freezing the weights of low-level layers and update weighs of high-level layers. 
With the limited availability of training data, stochastic gradient descent (SGD) can heavily be fluctuating the objective/loss function and hence overfitting can occur. To improve convergence and overcome overfitting, the mini-batch of stochastic gradient descent (\emph{mSGD}) was used to minimise the objective function, $E_{fine}(\cdot)$, with categorical cross-entropy loss  

\begin{eqnarray}
E_{fine}\left ( gl^{i},z(o^{j},\hat{W}) \right ) & = & -\sum_{i=1}^{c'k} gl^{i} \ln {z\left(o^{j},\hat{W}\right )},
\end{eqnarray}

where \( o^{j} \) is the set of input labelled images in the training, \( gl^{i} \) is the ground truth labels, while  $z\left(o^{j},\hat{W}\right )$ is the predicted output from a softmax function, where $\hat{W}$ is the converged weight matrix associated to the coarse transfer learning model.    

\subsection*{Performance evaluation}
In the downstream class-decomposition layer of \emph{4S-DT}, we divide each class within the image dataset into several sub-classes, where each subclass is treated as a new independent class. In the composition phase, those sub-classes are assembled back to produce the final prediction based on the original image dataset.
For performance evaluation, we adopted Accuracy (ACC), Specificity (SP) and Sensitivity (SN) metrics for multi-classes confusion matrix, the input image can be classified into one of ($c'$) non-overlapping classes. As a consequence, the confusion matrix would be a ($ N_{c'} \times N_{c'} $) matrix and the matrices are defined as:

    \begin{eqnarray} 
    {\rm Accuracy} (ACC) & = & \frac{1}{c'} \sum_{i=1}^{c'} \frac{TP_{i}+ TN_{i}}{TP_{i}+ TN_{i}+ FP_{i}+FN^{_{i}}}\,, \\ 
    {\rm Sensitivity } (SN) & = & \frac{1}{c'} \sum_{i=1}^{c'} \frac{TP_{i}}{TP_{i}+FN_{i}}\,,\\
    {\rm Specificity } (SP) & = & \frac{1}{c'} \sum_{i=1}^{c'} \frac{TN_{i}}{TN_{i}+FP_{i}}\,,    
    \end{eqnarray}

where $c'$ is the original number of classes in the dataset, $TP$ is the true positive in case of COVID-19 case and  $TN$ is the true negative in case of normal or other disease, while $FP$ and $FN$ are the incorrect model predictions for COVID-19 and other cases. Also, the $TP$, $TN$, $FP$ and $FN$ for a specific class $i$ are defined as:

\begin{eqnarray} 
    TP_{i}=\sum_{i=1}^{n}x_{ii}\\
    TN_{i}=\sum_{j=1}^{c}\sum_{k=1}^{c} x_{jk} ,j\neq i ,k\neq i\\
    FP_{i}=\sum_{j=1}^{c} x_{ji}   , j\neq i\\
    FN_{i}=\sum_{j=1}^{c} x_{ij}   , j\neq i,
\end{eqnarray}

where $x_{ii}$ is an element in the diagonal of the matrix. 
Having discussed and formalised the \emph{4S-DT} model in this section in detail, the following section validates the model experimentally. The model establishes the effectiveness of self-supervised super sample decomposition in detecting COVID-19 from chest X-ray images. 

\section{Experimental Results}
\label{results}

This section presents the datasets used in training and evaluating our \emph{4S-DT} model, and discusses the experimental results. 
\subsection{Datasets}

In this work, we used three datasets of labelled and unlabelled chest X-ray images, defined respectively as:
\begin{itemize}
\item \textbf{Unlabelled chest X-ray dataset}, a large set of chest X-ray images used as an unlabelled dataset: A set of 50,000 unlabelled chest X-ray images collected from three different datasets: 1) 336 cases with a manifestation of tuberculosis, and 326 normal cases from\cite{jaeger2013automatic, candemir2013lung}; 2) 5,863 chest X-Ray images with 2 categories: pneumonia and normal from \cite{kermany2018identifying}; and 3) a set of 43,475 chest X-ray images randomly selected from a total of 112,120 chest X-ray images, including 14 diseases, available from \cite{wang2017chestx}.

\item \textbf{COVID-19 dataset-A}, an imbalanced set of labelled chest X-ray with COVID-19 cases: 80 normal cases from \cite{6663723, 6616679}, and chest X-ray dataset from \cite{cohen2020covid}, which contains 105 and 11 cases of COVID-19 and SARS, respectively. We divided the dataset into two groups: 70$\%$ for training and 30$\%$ for testing. Due to the limited availability of training images, we applied different data augmentation techniques (such as: flipping up/down and right/left, translation and rotation using random five different angles) to generate more samples, see Table \ref{dataset1}.

\item \textbf{COVID-19 dataset-B}, we used a public chest X-ray dataset that already divided into two sets (training and testing), each set consists of three classes (e.g. COVID-19, Normal, and Pneumonia), see Table \ref{datasetB}. The dataset is available for download at: (\url{https://www.kaggle.com/prashant268/chest-xray-covid19-pneumonia}).


\end{itemize}

Note that chest X-ray images of dataset-A and datset-B are progressively updated and the distributions of images in these datasets (e.g. Tables \ref{dataset1} and \ref{datasetB}) can be considered as a snapshot at the time of submitting this paper. Therefore, any attempt to compare the performance of methods on such datasets at different points in time would be misleading. Moreover, the performance of the methods reported in this paper is expected to improve in the future with the growing availability of labelled images.

\begin{table}[]
\begin{center} 
\caption{The distribution of classes in COVID-19 dataset-A.} \label{dataset1} 
\begin{tabular}{ccccc}
\hline
Type & COVID-19 & SARS & Normal & Total \\
\hline
 Training  set &   74     &    8      &     56     &  138 \\
 Augmented training set &  662      &  69        &  504        &  1235 \\
 Testing  set 1 &  31    &     3      &   24    & 58  \\   
Testing set 2 &  283    &     30      &   216    & 529  \\   
\hline
\end{tabular}
\end{center} 
\end{table}

\begin{table}[]
\begin{center} 
\caption{The distribution of classes in COVID-19 dataset-B.}
\label{datasetB}
\begin{tabular}{cccc}
\hline
Class Name & Train & Test & Total \\
\hline
COVID-19    & 460   & 116  & 576   \\
Normal     & 1266  & 317  & 1583  \\
Pneumonia  & 3418  & 855  & 4273  \\
\hline
\end{tabular}
\end{center} 
\end{table}


All the experiments in our work have been carried out in MATLAB 2019a on a PC with the following configuration: 3.70 GHz Intel(R) Core(TM) i3-6100 Duo, NVIDIA Corporation with the donation of the Quadra P5000GPU, and 8.00 GB RAM.


\subsection{Self supervised training of \emph{4S-DT}}
\begin{figure}[hbt!]
    \centering
        \includegraphics[scale=0.25]{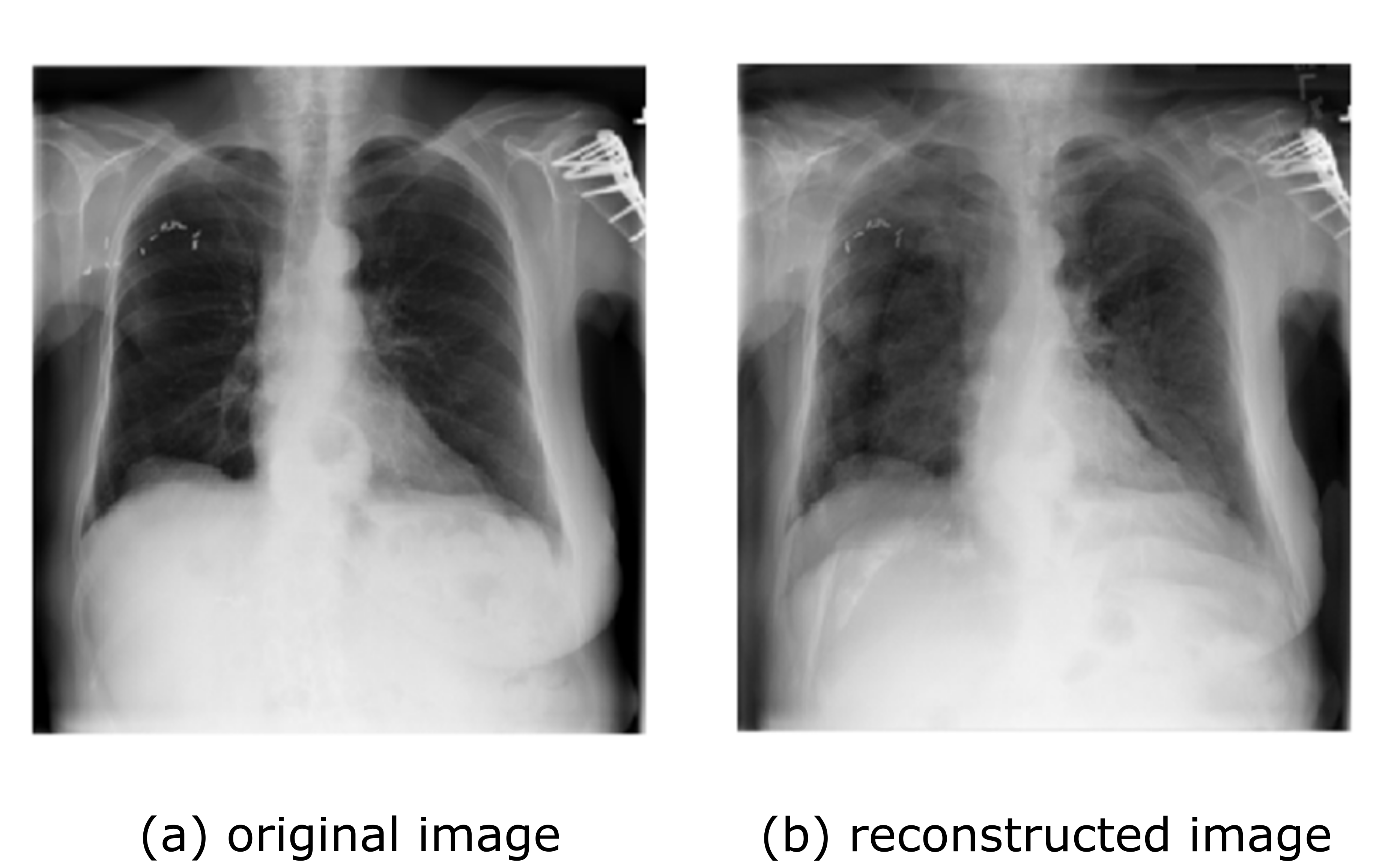}
        \caption{An example of a reconstructed chest X-ray image by our Autoencoder.}
        \label{ae}
    \end{figure}

\begin{figure}[hbt!]
    \centering
        \includegraphics[scale=0.32]{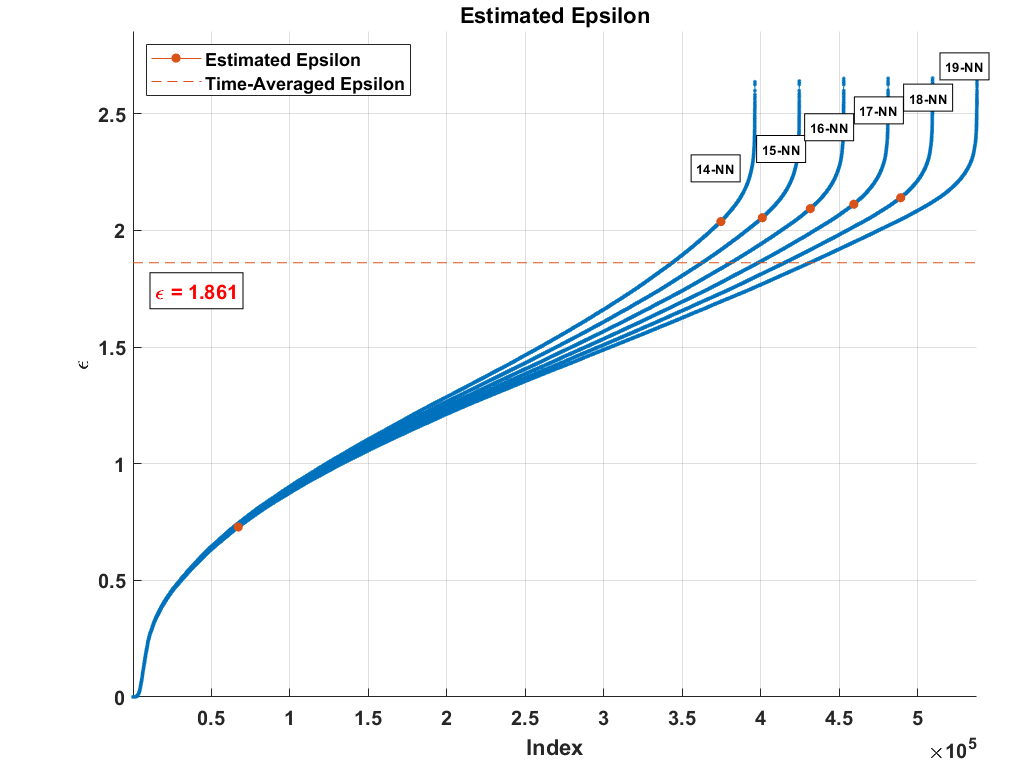}
        \caption{Estimation of the optimal $Eps$.}
        \label{epsilon}
    \end{figure}

We trained our autoencoder with 80 neurons in the first hidden layer and 50 neurons in the second hidden layer for the reconstruction of input unlabelled images, see Fig \ref{ae}. The trained autoencoder is then used to extract a set of deep features from the unlabelled chest X-ray images. The extracted features were fed into the \emph{DBSCAN} clustering algorithm for constructing the clusters (and hence the pseudo-labels). Since \emph{DBSCAN} is sensitive to the neighbourhood radius, we employed a k-nearest-neighbour (\emph{k-NN}) \cite{dudani1976distance} search to determine the optimal (\emph{Eps}) value. As demonstrated in Fig \ref{epsilon}, the optimal value for $Eps$ was 1.861. $MinPts$ parameter has been derived from the number of features ($d$) such that $MinPts \geq d+1$. Consequently, we used and tested different values for $MinPts$ parameter such as 51, 54, and 56 resulting in 13, 6, and 4 clusters respectively. For the coarse transfer learning, we used ResNet18 pre-trained \emph{CNN} model. The classification performance, on the pseudo-labelled samples, associated with the 13, 6, and 4 clusters were 48.1$\%$, 53.26$\%$, and 64.37$\%$, respectively. Therefore, we fix the number of clusters (and hence the number of pseudo labels) to be 4 in all experiments in this work.

\subsubsection{Downstream class-decomposition of \emph{4S-DT}}
We used AlexNet \cite{krizhevsky2012imagenet} pre-trained network based on a shallow learning mode to extract discriminative features of the labelled dataset. We set a value of 0.0001 for the learning rate, except the last fully connected layer (was 0.01), the min-batch size was 128 with the minimum of 256 epochs, 0.001 was set for the weight decay to prevent the overfitting through training the model, and 0.9 for the momentum speed. At this stage, 4096 attributes were obtained, therefore we used \emph{PCA} to reduce the dimension of feature space. 
For the class decomposition step, we used $k$-means clustering \cite{wu2008top}, where $k$ has been selected to be 2 and hence each class in  \( \mathbf{L} \)  has been further divided into two subclasses, resulting in a new dataset with six classes. The adoption of $k$-means for class decomposition with $k=2$ is based on the results achieved by the DeTraC model in \cite{abbas2020detrac}. 

\subsection{Classification performance on COVID-19 dataset-A}


We first validate the performance of \emph{4S-DT} with ResNet18 (as the backbone network) on the 58 test images (i.e. testing set 1), where augmented training set is used for training, see Table \ref{dataset1}. Our ResNet architecture consists of residual blocks and each block has two $3\times3$ $Conv$ layers, where each layer is followed by batch normalisation and a ReLU activation function. 
Our ResNet architecture consists of residual blocks and each block has two $3\times3$ $Conv$ layers, where each layer is followed by batch normalisation and a ReLU activation function. Table~\ref{descrResNet} illustrates the adopted architecture used in our experiment. 

\begin{figure}[hbt!]
    \centering
        \includegraphics[scale=0.18]{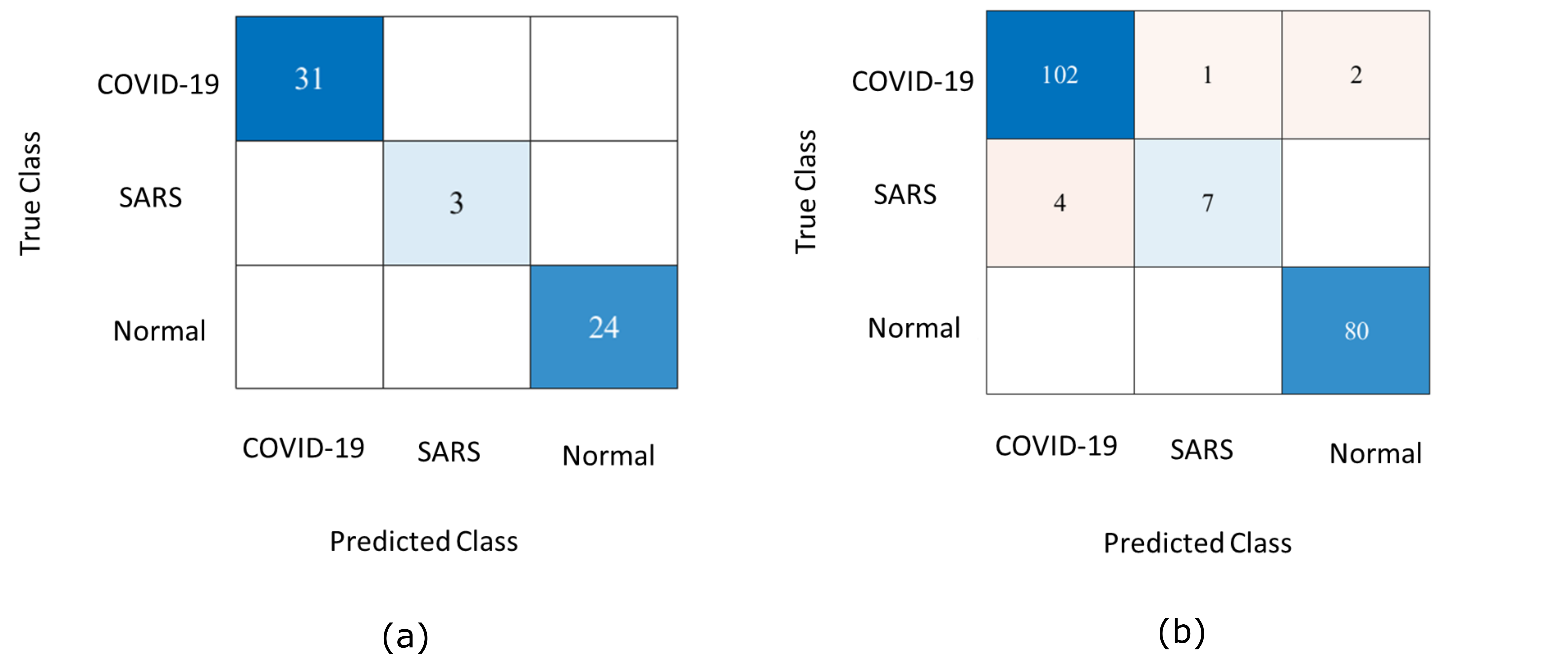}
        \caption{Confusion matrix obtained by a) \emph{4S-DT} on the 58 test set, and b) \emph{4S-DT} when trained on augmented images only and tested on the 196 cases.}
        \label{CM}
    \end{figure}

During the training of the backbone network, the learning rate for all the \emph{CNN} layers was fixed to 0.0001 except for the last fully connected layer (was 0.01) to accelerate the learning. The mini-batch size was 256 with a minimum of 200 epochs, 0.0001 was set for the weight decay to prevent the overfitting through training the model, and the momentum value was 0.95. The schedule of drop learning rate was set to 0.95 every 5 epochs. The results were summarised in Table~\ref{FinResNet}. 
Moreover, we also compare the performance of the proposed model without the self supervised sample decomposition component (i.e. w/o \emph{4S-D} or \emph{DeTraC-ResNet18} \cite{abbas2020detrac}) and without both \emph{4S-D} and class-decomposition (w/o \emph{4S-D}+\emph{CD} or ResNet18 \cite{he2016deep} pre-trained network on) the 58 testing set. \emph{4S-DT} has achieved 100$\%$ accuracy in the detection of COVID-19 cases with 100$\%$ (95$\%$ confidence interval (\emph{CI}): 96.4$\%$, 98.7$\%$) for sensitivity and specificity (95$\%$ \emph{CI}: 94.5$\%$, 100$\%$), see Fig. \ref{CM}. As illustrated by Fig. \ref{CM} and Table \ref{Test58}, \emph{4S-DT} shows a superiority and a significant contribution in improving the transfer learning process with both the self supervised sample decomposition and downstream class-decomposition components. Also, we applied \emph{4S-DT} based on ResNet18 pre-trained network on the original classes of COVID-19 dataset with an imbalance classes after eliminating the samples from the training set. As we see in Fig. \ref{CM}, \emph{4S-DT} has achieved 96.43$\%$ accuracy (95$\%$ CI: 92.5$\%$, 98.6$\%$) in the detection of COVID-19 cases with sensitivity 97.1$\%$ (95$\%$ \emph{CI}: 92.24$\%$, 97.76$\%$) and 95.60 $\%$ (95$\%$ \emph{CI}: 93.41$\%$, 96.5$\%$) for specificity. 
    
\begin{table}[]
\begin{center} 
\caption{The adopted ResNet architecture used in the fine-tuning study in our experiments.} 
\label{descrResNet} 
\begin{tabular}{ccc}
\hline
Layer Name    & ResNet\_18      & Output Size    \\
\hline
Conv1    & $7\times 7$ , 64, stride (2)                                     & $112\times112 \times 64$ \\
\hline
Conv2  & \begin{tabular}[c]{@{}c@{}}Layer-Res2a\\ Layer-Res2b\end{tabular} & $56\times56\times 64$   \\
\hline
Conv3   & \begin{tabular}[c]{@{}c@{}}Layer-Res3a\\ Layer-Res3b\end{tabular} & $28\times28 \times 128$  \\
\hline
Conv4   & \begin{tabular}[c]{@{}c@{}}Layer-Res4a\\
Layer-Res4b\end{tabular} & $14 \times14 \times256$     \\
\hline
Conv5   & \begin{tabular}[c]{@{}c@{}}Layer-Res5a\\ Layer-Res5b\end{tabular} & $7\times 7 \times512$        \\
 \hline
\end{tabular}
\end{center} 
\end{table}


\begin{table}
\begin{center} 
\caption{Classification measurements obtained by \emph{4S-DT} without both \emph{4S-D} and class-decomposition (w/o \emph{4S-D}+\emph{CD}) and without \emph{4S-D} (w/o \emph{4S-D}) only on the 58 labelled chest X-ray images (i.e. testing set 1).}
\label{Test58} 
\begin{tabular}{ccc ccc}
\hline
\multicolumn {3}{c}{(w/o \emph{4S-D}+\emph{CD})}&\multicolumn{3}{c}{(w/o \emph{4S-D})}  \\
\hline
ACC ($\%$) & $SN$ ($\%$) & $SP$ ($\%$) &  ACC ($\%$) & $SN$ ($\%$) & $SP$ ($\%$)\\ 
\hline
89.66\%& 96.77\% & 81.48\% &  93.10\%  &                            96.77 \%   & 88.89 \% \\
\hline
\end{tabular}
\end{center} 
\end{table}

To allow for further investigation and make testing of COVID-19 detection more challenging, we applied the same data augmentation techniques (used for the training samples) to the small testing set to increase the number of testing samples. The new test sample distribution, we called testing set 2, contains 283 COVID-19 images, 30 SARS images, and 216 normal images, see Table \ref{dataset1}. Consequently, we used testing set 2 for testing and augmented training set for training (see Table \ref{dataset1}), unless otherwise mentioned, for the performance evaluation of all methods in the experiments described below. We validated the performance of a) the full version of \emph{4S-DT} with \emph{4S-D} component and b) without \emph{4S-D}. For a fair comparison, we used the same backbone network (i.e. ResNet18) with the downstream class-decomposition component, where both versions have been trained in a shallow and fine-tuning mode. As illustrated by Table \ref{FinResNet}, \emph{4S-D} component shows significant improvement in shallow- and fine-tuning modes in all cases. More importantly, our full version model with \emph{4S-D} demonstrates better performance, in each case, with less number of epochs, confirming its efficiency and robustness at the same time.

\begin{table*}
\begin{center} 
\caption{Overall classification performance of \emph{4S-DT} model with and without self-supervised sample decomposition (\emph{4S-D}) component, on testing set 2.}
\label{FinResNet} 
{
\begin{tabular}{c c  c  c  c c c c c} 
\hline 
Layer& \multicolumn {4}{c} {without \emph{4S-D} } &\multicolumn {4}{c} {with \emph{4S-D}}\\ 
& ACC & SN & SP &Epochs&  ACC  & SN & SP & Epochs\\ 
& ($\%$) & ($\%$) & ($\%$) &$\#$&  ($\%$) &  ($\%$) &  ($\%$) &$\#$ \\
\hline 
Shallow  &     92.12& 64.13  & 94.2&  61&
                      97.48& 88.64  & 98.01& 29 \\
Res5b&  93.84& 64.18  & 94.06& 75&     
                      97.66& 93.08  & 98.41& 42 \\
Res5a&  93.84& 64.18  & 94.06& 83&   
                      97.23& 87.33    & 97.73& 33 \\
Res4b&  93.96& 64.33  &94.20&112& 
                      96.8&     86.3   &97.91&32 \\
Res4a&  94.04& 64.52    &94.37&82 &    
            97.99 & 87.15& 98.10&25\\
Res3b&  94.34& 64.16  &94.25& 73&                               97.99& 87.71  &97.42&32 \\

\hline 
\end{tabular} 
} 
\end{center} 
\end{table*}

\begin{figure}[hbt!]
\centering
\includegraphics[scale=0.5]{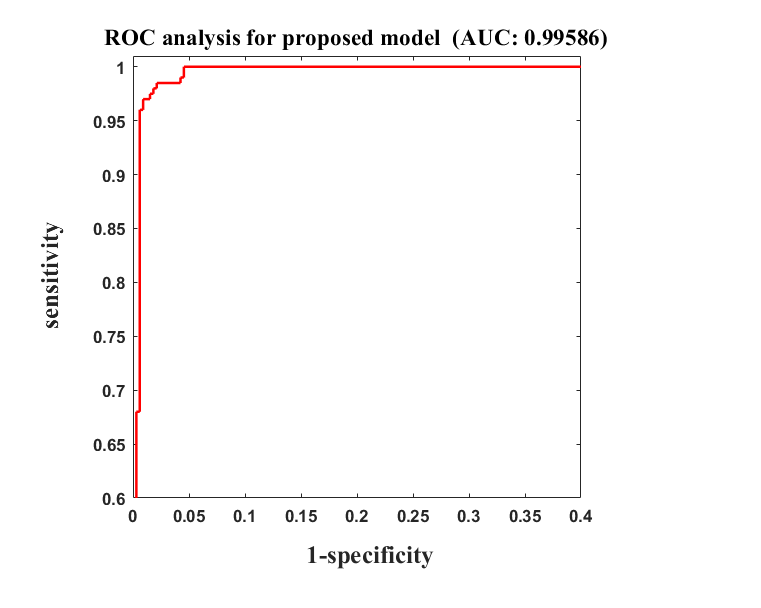}
\caption {ROC curve obtained during the training of \emph{4S-DT} with ResNet pre-trained model.}
\label{ROC}
\end{figure}

Moreover, we compared the classification performance of \emph{4S-DT} with other models used for COVID-19 detection, including GoogleNet \cite{szegedy2015going}, \emph{DeTraC}. \emph{4S-DT} has achieved a high accuracy of $97.54\%$ (95$\%$ CI: 96.22\%, 98.91\%) with a specificity of $97.15\%$ (95$\%$ CI: 94.23$\%$, 98.85$\%$) and sensitivity of $97.88\%$ (95$\%$ CI: 95.46$\%$, 99.22$\%$) on the 529 test chest X-ray images of test set 2, see Table \ref{tabshallow}. Moreover, as shown by Table \ref{tabshallow}, \emph{4S-DT} has demonstrated superiority in performance, confirming its effectiveness in improving the classification accuracy of transfer learning models. Finally, Fig. \ref{ROC} shows the Area Under the receiver curve (AUC) between the true positive rate and false positive rate obtained by \emph{4S-DT}, with AUC value of 99.58$\%$ (95$\%$ CI: 99.01$\%$, 99.95$\%$), to confirm its robustness behaviours during the training process. 

\begin{table} 
\begin{center} 
\caption{Classification performance of \emph{4S-DT} and other models on testing set 2 of the COVID-19 dataset.} 
\label{tabshallow} 
{
\begin{tabular}{c c c c } 
\hline 
Model& ACC ($\%$)& SN ($\%$)& SP ($\%$)\\ 
\hline 
\emph{4S-DT} (ResNet)	& 97.54& 97.88 & 97.15\\
\emph{4S-DT} (GoogleNet)	&94.15&97.07&93.08\\
\emph{4S-DT} (Vgg19)	&95.28&93.66&97.15\\
\emph{DeTraC- ResNet18} \cite{abbas2020classification}	&95.12&97.91&91.87\\
ResNet18 \cite{he2016deep} &92.5&65.01&94.3\\
\emph{DeTraC-GoogleNet} \cite{abbas2020classification}&91.01&76.03&82.6\\
\emph{DeTraC-Vgg19} \cite{abbas2020classification}&93.42& 89.71 & 95.7\\
\hline 
\end{tabular} 
} 
\end{center} 
\end{table}

\subsection{Classification performance on COVID-19 dataset-B}
\begin{figure*}[hbt!]
    \centering
        \includegraphics[scale=0.45]{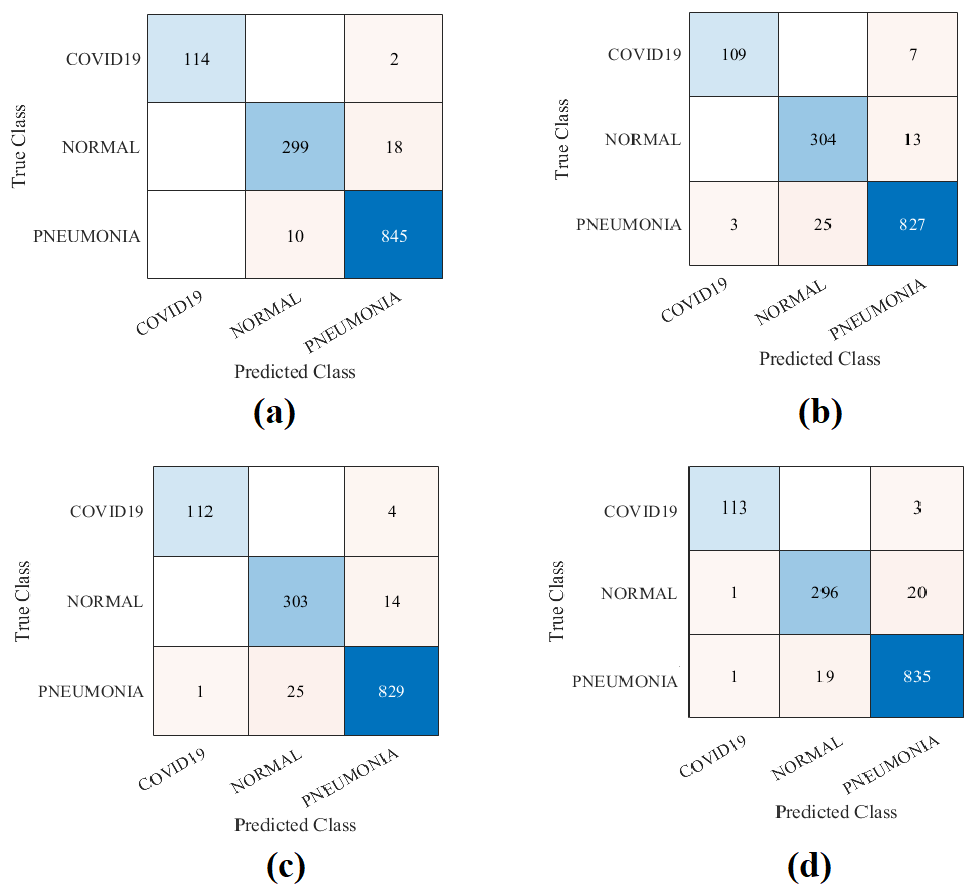}
        \caption{The confusion matrix results of COVID-19 dataset-B obtained by \emph{4S-DT}, based on different pre-trained networks: a) VGG19, b) GoogleNet, c) ResNet and d) Mobilenetv2.}
        \label{CM2}
    \end{figure*}

To evaluate the performance of \emph{4S-DT} on COVID-19 dataset-B, we applied different ImageNet pre-trained CNN networks such as: VGG19 \cite{simonyan2014very}, ResNet \cite{szegedy2017inception}, GoogleNet \cite{szegedy2015going}, and Mobilenetv2 \cite{sandler2018mobilenetv2}. Parameter settings for each pre-trained model during the training process are reported in Table \ref{parameSett}.
Transfer learning has been accomplished via deep tuning scenario (with 15 epochs and SGD was the optimiser). The classification performance on COVID19 cases was reported in Table \ref{Newdata}. Fig \ref{CM2} illustrates the confusion matrix obtained by each pre-trained Networks for each class in the dataset. As demonstrated by Table \ref{Newdata}, \emph{4S-DT} has achieved a high accuracy of $99.8\%$ (95$\%$ CI: 99.44 \%, 99.98\%), with sensitivity of 99.3\%$ (95$\%$ \emph{CI}: 93.91$\%$, 99.79$\%$), and specificity of 100\%$ (95$\%$ \emph{CI}: 99.69$\%$, 100$\%$) in the detection of COVID-19 cases.  

\begin{table*}[]
\begin{center} 
\caption{Parameters settings for each pre-trained model used for training \emph{4S-DT} model on COVID-19 dataset-B.}
\label{parameSett}
\begin{tabular}{ccccc}
\hline 
Pre-trained model & Learning rate & MB-Size & Weight decay & Learning rate-decay \\
\hline 
VGG19     &   0.01 &   32 &  0.0001 &  0.9 every 2 epochs        \\
GoogleNet &  0.0001 & 128   & 0.001  & 0.95 every 2 epochs   \\ 
ResNet    &  0.001   & 256   & 0.0001  &  0.9 every 3 epochs        \\
Mobilenetv2 &    0.001   & 64  & 0.001 & 0.9 every 2 epochs\\                    
\hline  
\end{tabular}
\end{center} 
\end{table*}

\begin{table}[]
\begin{center}
\caption{The classification performance of COVID-19 dataset-B obtained by \emph{4S-DT} based on different pre-trained models.} 
\label{Newdata}
\begin{tabular}{cccc}
\hline
pre-trained model & Acc (\%) & SN(\%) & SP(\%) \\
\hline
VGG19        & 99.8 & 99.7 & 100   \\
GoogleNet    & 99.2     &93.9    & 99.7     \\
ResNet       & 99.6     & 96.5   & 99.9    \\
Mobilenetv2    & 99.6     & 97.4   & 99.8       \\
\hline
\end{tabular}
\end{center}
\end{table}

\section{Discussion and conclusion}
\label{discussion}

The diagnosis of COVID-19 is associated with the pneumonia-like symptoms that can be revealed by genetic and imaging tests. Chest X-ray imaging test provides a promising fast detection of COVID-19 cases and consequently can contribute to controlling the spread of the virus. In medical image classification, paramount progress has been made using ImageNet pre-trained convolutional neural networks (\emph{CNN}s), exploiting the high availability of large-scale annotated image datasets. The historical conception of such approaches has been comprehensively explored through several transfer learning strategies, including fine-tuning and deep-tuning mechanisms. They usually require an enormous number of balanced annotated images distributed over several classes/diseases (which is impractical in the medical imaging domain). In medical image analysis, data irregularities still remain a challenging problem, especially with the limited availability of confirmed samples with some diseases such as COVID-19 and SARS, which usually results in miscalibration between the different classes in the dataset. Consequently, COVID-19 detection from chest X-ray images presents a challenging problem due to the irregularities and the limited availability of annotated cases. 

Here, we propose a new \emph{CNN} model, we called Self Supervised Super Sample Decomposition for Transfer learning (\emph{4S-DT}) model. \emph{4S-DT} has been designed to cope with such challenging problems by adapting a self-supervised sample decomposition approach to generate pseudo-labels for the classification of unlabelled chest X-ray images as a pretext learning task. \emph{4S-DT} has also the ability to deal with data irregularities by a class-decomposition adapted in its downstream learning component. \emph{4S-DT} has demonstrated its effectiveness and efficiency in coping with the detection of COVID-19 cases in a dataset with irregularities in its distribution. In this work, we used 50,000 unlabelled chest X-ray images for the development of our self-supervised sample decomposition approach to perform transfer learning with an application to COVID-19 detection. We achieved an accuracy of $97.54\%$ with a specificity of $97.15\%$ and sensitivity of $97.88\%$ on 529 test chest X-ray images (of COVID-19 dataset-A), i.e. testing set 2, with 283 COVID-19 samples. We also achieved a high accuracy of $99.8\%$ in the detection of COVID-19 cases of COVID-19 dataset-B.   
 
With the continuous collection of data, we aim in the future to extend the development and validation of \emph{4S-DT} with multi-modality datasets, including clinical records. As a future development, we also aim to add an explainability component to increase the trustworthiness and usability of \emph{4S-DT}. Finally, one can use model pruning and quantisation to improve the efficiency of \emph{4S-DT}, allowing deployment on handheld devices.

	\bibliographystyle{IEEEtran} 
\bibliography{ref}



\end{document}